# Microfluidics in Late Adolescence

*George Whitesides - Harvard University*

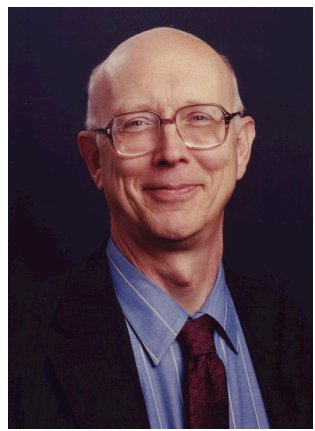

### Biography

*George M. Whitesides. Woodford L. and Ann A. Flowers University Professor. Born, 1939, Louisville, KY. A.B., Harvard, 1960. Ph.D., 1964, California Institute of Technology (with J.D. Roberts). Faculty: Massachusetts Institute of Technology, 1963 to 1982; Harvard University, 1982-present.*

*Memberships and Fellowships. American Academy of Arts and Sciences, National Academy of Sciences, National Academy of Engineering, American Philosophical Society, Fellow of the American Association for the Advancement of Science, Fellow of the Institute of Physics, New York Academy of Sciences, World Technology Network, Honorary Member of the Materials Research Society in India, Royal Netherlands Academy of Arts and Sciences, Foreign Fellow of the Indian National Science Academy, and Honorary Fellow of the Royal Society of Chemistry (UK).*

*Present research interests include: physical and organic chemistry, materials science, biophysics, complexity and emergence, surface science, microfluidics, optics, self-assembly, science for developing economies, catalysis, origin of life, rational drug design, cell-surface biochemistry, simplicity.*

### Stimuli

**Analytical Methodology.** The original stimulus for the development of what is now called "microfluidics" was the desire to reduce the size and complexity of (bio)chemical analytical system. A broad objective – a "Lab on a Chip" – was a term famously attributed to Andreas Manz [1], but the broader expectation was for a microtechnology for handling fluids (especially biological fluids) that would become as ubiquitous as "microelectronics" was in handling another fluid (electrons).

Four specific themes provided the stimuli for the initial development of the subject. i) **National Security**. The initial perceived technical need was to develop portable analytical systems for use in chemical and biological defense. This early work (largely in the 1990s) was funded by DARPA, and jump-started the field, as it has developed. ii) **Pharmaceutical and Clinical Analysis**. Another hoped-for application was to improve the efficiency of lead (and perhaps drug) discovery and development in the pharmaceutical industry. In this period, the pharmaceutical industry was unhappily contemplating the high cost and inefficiency of discovery of new drugs, and betting heavily on high-throughput screening, and related technologies. iii) **Genomics**. The explosive development of commercial devices for sequencing nucleic acids provided a third requirement for methods for manipulating small quantities of liquids. iv) **Point-of-Use Microanalysis**. Public health, environmental monitoring, health care for the developing world, agricultural chemistry, and medical analysis at the point of care (PoC) all provided further potential applications of small, affordable, simple-to-use bioanalytical systems.

## Early History

"Microfluidics" developed along two, weakly communicating tracks. One was largely academic, and focused on technology for manipulating fluids in microchannels. The second—and often preexisting—one was industrial, and developed the fluid-management systems for ink-jet printing, analytical instrumentation, lubricated bearings, spray painting, and many others.

**Silicon and Glass.** An interesting early competition between technologies for use in microfluidics involved silicon [2] and glass [3] on the one hand, and organic polymers on the other. The argument for the former was that the technology required to fabricate microchannels was already highly developed (from microelectronics), and there was no need to develop anything new. The arguments for the latter were many, and included simplicity, use of transparent materials, ease of prototyping, and others. In the event, the needs of microfluidics and microelectronics are almost entirely different, and most of the field has developed using polymer-based systems.

**PDMS**. By far the most important material for microfluidics has been poly(dimethyl siloxane), developed independently by us [4] and by Manz [5]. This material – when combined with the extraordinarily convenient methods that developed for its fabrication – made the design of even quite complicated microfluidic systems sufficiently simple it became the standard for exploratory work. Whether PDMS will ultimately be used for large-volume production (rather than structural polymers such as polycarbonate) remains to be established, but for academic research, and for prototyping, it has been unbeatable.

**The Glucometer.** A very important, if specialized, microfluidic electrochemical device had, in fact, been developed much earlier by Adam Heller [6], and provided a proof-of-concept for microfluidics. This system–which presaged the current interest in paper-based electrochemical devices–was a microfluidic system that allowed the analysis of glucose in blood from a finger-stick. It is certainly the most practically important diagnostic microanalytical system, and it is curious that PDMS microfluidics developed completely independently of this very successful, earlier paper-based microfluidic system.

## Some Key Steps

**Systems of Channels.** A combination of ink-jet printing and 1:1 contact photolithography (to make masters), casting, and simple sealing made it possible to make even complex microfluidic systems with channel widths of 10-1000 microns with little effort and expense [7]. The dimensions are, of course, much larger than those in microelectronic devices, but they are correct for the sample volumes of interest in bioanalysis, in handling blood, and in cell biology. PDMS is transparent, non-toxic, biocompatible, and relatively inexpensive.

**Laminar Flow; from Large to Small Channels.** Understanding the physics of flow of liquids in microchannels was made much easier by the fact that the descriptions developed in chemical engineering for flows in large pipes scaled directly to microchannels [8-10], and the understanding of these relationships allowed intuitive design of new systems.

**Cell Biology.** Early experiments demonstrated that PDMS was fully compatible with the growth of mammalian cells [11], and with the culture of microorganisms and simple multicellular organisms (e.g., C. elegans [12]). The techniques used in the earliest experiments are remarkably close to those being reexamined now for technologies such as "organ on a chip."

**"Quake Valves."** One of the limitations of early-generation PDMS-based devices (and one that indeed remains to this day) is the absence of flexible, small valves and switches to allow the control of fluid flows in microfluidic systems. Quake developed pneumatic valves that served as effective valves for fluidic microsystems [13], and allowed the construction of very complex systems. We [14] and Beebe et al [15] also developed microfluidic valves based on prefabricated structures or hydrogel, respectively. The Quake valves have been the basis for important developments in digital biology, and have produced important concepts for synthesis of short-lived 19F labeled drugs for PET imaging. The weakness of these valves is that although they are small, and the microfluidic system remains small, the associated computers, pressure sources, solenoid valves, and connectors are not small, and the overall *system* remains large. There is still no equivalent of a transistor for fluidic microsystems.

**Microsystems Involving Fluids: Microwells, Fluidic Systems Internal to More Complex Devices.** While PDMS-based microfluidics developed into an independent field, with its own journals (e.g., Lab on a Chip) and technical foci, microfluidic systems taken from a range of technologies were also being embedded (often invisibly) in a number of other technologies. Examples are the robotic fluidic systems used in high-throughput screening and clinical diagnostics, the microsystems used in in-line chemical synthesis, the internal fluid-handling systems for the instruments used in genomic sequencing and microchemical analysis, the inlets (and separations systems) used in CE/MS and LC/MS, the flow focusing and switching systems developed in fluorescence-activated cell sorting (FACS), and others.

## Some Newer Developments and Applications

The field continues to produce new methodologies. Four represent examples. i) **Droplets and Multiphase Systems.** These studies started with examination of gas bubbles by Gañán-Calvo *et al* [16], in aqueous fluids by Quake *et al* [17], and by us, in bubble rafts and trains (with the demonstration of remarkably complex behaviors [18]), and then (Weitz *et al* [19]) of multiple-drop systems. These are beginning to be applied in areas such as complexity and infochemistry (Whitesides [20,21]), digital biology (Ismagilov *et al* [22]), and single-cell genomics (Weitz *et al* [23] and Mathies *et al* [24]). ii) **"Organs on a Chip."** Growth of cells in shaped channels, especially with mechanical stimulation (a capability that is uniquely enabled by PDMS and other elastomers, and may mimic aspects of tissues) [25], is a possible basis for moving beyond simple cell-culture in microchannels to structured tissues that in some ways mimic the behavior of cells in tissues (e.g., lung, heart, and colon). iii) **"Slip-chip"** [26] is a methodology for carrying out serial dilutions and mixing that provides a simple entry into certain analyses, and for digital biology. iv) **Electrowetting** [27] moves aqueous fluids on flat surfaces – unconstrained by a microchannel – using electric fluids generated locally. v) **Microwell Systems.** The distribution of fluids in microwells is, again, distinct from "PDMS microfluidics," but in the hands of Walt *et al* [28] (also the originator of Illumina Inc – the company that has provided an important part of the foundation of modern nucleic acid analysis) it is now offering an exceptional opportunity for studying the catalytic activity of single enzymes.

## Our Current Work

**Simple Functions.** Our own recent work has emphasized the development of microfluidic systems that are simple in function. The immediate intent was to provide technology for medical diagnosis in the developing world (although often complex in mechanism) [29].

**Micro Paper Analytical Devices.** PDMS devices – although simple relative to clinical

analyzers – are probably still too expensive and complicated to be used in resource-constrained environments. We have developed "paper diagnostics" [30] – a broadly applicable method of fabricating microfluidic devices by patterning channels by printing hydrophobic waxes in hydrophilic paper to form channels. These systems are, in fact, microchannel or microfluidic systems, in which the channels are the spaces between the fibers of cellulose. They require no power to move the fluids (capillarity serves), and can perform many of the standard types of bioassays (metabolites and toxins, ELISA, detection of nucleic acid sequences, and others). Although the original systems were designed to be read by eye (colorimetrically), it is now evident that most PoC diagnostics – in both the developed and developing worlds – will be read using apps and simple readers attached to cellphones, and the current generation of paper diagnostic devices is designed to be compatible with that form of readout. As a technology, paper diagnostics are very successful, and are now beginning to move through regulatory clearance.

**Paper-Based Electrochemistry.** For a number of reasons, paper-based systems are ideally matched with electrochemistry: the paper provides a hydrophilic matrix that holds an aqueous fluid, that inhibits convective movement of fluid, and that can be channeled and timed precisely. These methods are, in a sense, the descendants of the Glucometer, although they evolved independently (starting with experiments by Henry [31], Crooks [32], us [33], and others). **Soft Robotics.** An interesting off-shoot of microfluids has now appeared – improbably – in robotics. The same methods of fabrication developed for microfluidic systems (with, of course, modifications in details of design), when inflated with air, or evacuated, are excellent soft actuators, and have become the basis of a very rapidly growing new field: "soft robotics." We produced the first demonstration of this extension of microfluidics in 2011[34], and multiple companies are already producing commercial actuators. These systems are important for their ability to carry out "collaborative" tasks: that is, to work safely in proximity with (and even in contact with) humans.

## Futures

What is now the future of microfluidics? My *opinion* is that it is now past the initial phase of "technology developments:" it has developed at least the first level of the methodologies required to become a large-scale technology. It is a key support for both chemical and biochemical analysis, and thus for biology. The science, and initial prototyping of applications, are successful; now comes the expensive tasks in engineering, production, and market development. Of these, I list only five. i) **Public Health (in both Developed and Developing Regions), Point-of-Care Analysis, Environmental Testing, and National Security.** All of these subjects require functionally simple analytical methods that can be used remotely from central laboratories. ii) **Drug Development.** The pharmaceutical industry *must* become more efficient in generating new drugs. More "human-like" testing (probably based on cell and organ culture) is one of the most promising approaches to this problem. Cells and organs on chips also provide new approaches to fundamental cell biology. iii) **The Web.** The web has opened entirely new approaches (unlimited data, deep learning, neural networks, and other forms of artificial intelligence) to understanding, monitoring, and testing large populations of patients, and discovering patterns in large numbers of data. Although some of this information can be obtained non-invasively, ultimately many tests will require access to information present in biological fluids such as blood and urine. Simple methods of sampling and analyzing fluids will also become more, rather than less, important as precision medicine is increasingly mandated for patient safety and management of costs. iv) **Fluids in Biology.** All of life involves moving fluids. From cytoplasmic streaming in single cells to the complex fluidic systems that process blood, lymph, urine, and other fluids, complex fluids are everywhere. We barely understand the physical – much less the biophysical – properties of these fluids, and microfluidic systems will be required to study them. v) **Organic Synthesis.** Even this highly-developed field may be revolutionized by a combination of microfluidic synthesis and (finally) computer-based path selection.